\documentclass{article}

\usepackage{amssymb}
\usepackage{amsmath}
\usepackage{amscd}
\usepackage[dvips]{epsfig}

\def\hs{\hspace{1cm}}

\begin{document}

\begin{titlepage}
\noindent
hep-th/9912016
\hfill ITP--UH--20/99 \\

\vskip 2.0cm

\begin{center}

{\large\bf WARD IDENTITIES AND THE}\\

\medskip

{\large\bf  VANISHING THEOREM FOR LOOP AMPLITUDES}\\

\medskip

{\large\bf  OF  THE CLOSED N=2 STRING}

\vskip 1.5cm

{\large Klaus J\"unemann \quad and \quad Olaf Lechtenfeld}

\vskip 0.5cm

{\it Institut f\"ur Theoretische Physik, Universit\"at Hannover}\\
{\it Appelstra\ss{}e 2, 30167 Hannover, Germany}\\
{E-mail: junemann, lechtenf@itp.uni-hannover.de}\\
 
\end{center}
\vskip 2.5cm

\begin{abstract}
The existence of a ground ring of ghost number zero operators 
in the chiral BRST cohomology of the $N{=}2$ string is used 
to derive an infinite set of Ward identities for the closed-string 
scattering amplitudes at arbitrary genus. These identities are 
sufficient to  rederive the well known vanishing theorem 
for loop amplitudes with more than three external legs. 
\end{abstract}

\vfill
\end{titlepage}

\section{Introduction}

The explicit computation of loop amplitudes in string theory 
is notoriously difficult. Even for the technically most simple
theory -- the bosonic string -- the level of mathematical complexity 
is impressing if one tries to go beyond one loop. 
Adding fermions and supersymmetry on the world-sheet does not improve the 
situation. On the contrary, the calculations become still more intricate
and only a few explicit results exist. 
It seems that even the general formalism has not yet been fully 
worked out \cite{Pol}.

Fortunately, explicit computations can sometimes be replaced by more 
indirect methods, often related to symmetry arguments.  
It is thus not surprising that 
for the  $N{=}2$ string (i.e. the theory based on extended supersymmetry on
the world-sheet; see \cite{M} for a general 
review and \cite{BL2} for a discussion of loop amplitudes) 
Berkovits and Vafa succeeded  to avoid the 
evaluation of the path integral and obtained powerful results for 
loop amplitudes by embedding the theory into an $N{=}4$ topological theory
\cite{BV}.
In fact, they found that all amplitudes with more than three external legs 
vanish to all orders in the loop expansion. 
The purpose of this letter is to give an alternative derivation of this 
result. Our approach has the advantage 
 that conceptually it is very clear what is going on since the equations
used to derive the vanishing of the amplitudes can nicely be interpreted 
as Ward identities of an infinite set of unbroken symmetries in target space. 
Another interesting point is that from a technical point of view 
our analysis rests on the picture dependence of the BRST cohomology of the 
$N{=}2$ string at zero momentum and   demonstrates what kind of information
may be stored  in the still somewhat obscure picture phenomenon \cite{FMS}. 
Maybe, this lesson can also be useful in some way for the $N{=}1$ string.

The  letter is organised as follows: In the next section 
we recall some facts about the BRST cohomology of the $N{=}2$ string. 
These results will  be used in section three to derive an infinite set of 
target space Ward identities which will then be explicitly 
evaluated so that the vanishing of the loop amplitudes directly follows. 
We conclude with some further remarks  and a brief discussion of the 
reliability of our arguments.

\section{Symmetries and ground ring of the $N{=}2$ string}
One of the attractive features of the BRST approach to closed string theory 
is that it provides an efficient means to analyse symmetries 
in target space. More precisely,  unbroken target space symmetries 
generally\footnote{There are exceptions, see section five of 
\cite{BD}.}  lead to the existence of ghost number one cohomology classes 
(in conventions where  physical states have ghost number two).
A detailed explanation of this fact is given in \cite{WZ} (see also 
\cite{BBH}) where in addition an elegant method to derive the corresponding 
Ward identities -- briefly reviewed below -- is described. 

Due to the fact that the closed string Fock space factorises into right- 
and left-moving parts ghost number one cohomology classes can be further 
characterised: they are most conveniently constructed as a product 
of a holomorphic piece of ghost number zero and an antiholomorphic piece 
of ghost number one. The latter is usually  the right-moving part of 
a physical vertex operator, taken at some discrete value of the 
momentum whereas the former very often is  just  the unit operator.
If, however, the chiral (= left-moving) cohomology at ghost number zero 
contains further elements besides the unit operator more closed string 
operators of ghost number one can be constructed  resulting in a 
much richer symmetry structure. An example is the bosonic 
string in two dimensions \cite{GM}. Moreover, interesting algebraic 
structures emerge. The BRST cohomology possesses a natural 
multiplication rule, additive in ghost number. 
The ghost number zero cohomology therefore forms 
a ring under this multiplication (the so-called ground ring). As has been 
emphasised in \cite{W} the structure constants of this ground ring 
encode much information about the symmetry of the theory\footnote{There 
exist two further operations --  the Gerstenhaber-bracket 
and the $\Delta$ operation  -- which, together with the ring multiplication, 
give the BRST cohomology the structure of a BV-algebra \cite{LZ}.}.  

The $N{=}2$ string has been studied along these lines in \cite{Gi,Li,JLP}. 
Based on the fact that so many of its scattering amplitudes are 
known or conjectured to vanish \cite{BV,OV,hipp} and comparison with 
the field theory that reproduces  tree-level scattering \cite{OV,PBR} 
it seemed very plausible that in this theory a  large symmetry group is 
realized. In fact,
a ground ring of the $N{=}2$ string has recently been found  in 
\cite{JLP} and will now briefly be reviewed.
The construction looks somewhat unconventional because 
it does not restrict to operators of  a single picture only, but 
takes into account the full picture degeneracy of the Fock 
space\footnote{This 
construction is non-trivial due to the picture dependence 
of the BRST cohomology of the $N{=}2$ string at zero momentum \cite{JL}.}. 
However, starting from this ground ring one may derive powerful 
Ward identities as has been shown for tree amplitudes in  \cite{JLP}
and will be demonstrated in this letter for loop amplitudes. 

At zero ghost number chiral  cohomology classes 
occur only for vanishing 
momentum. For low-lying picture numbers and ghost number zero 
the cohomology problem is rather 
straightforward to solve\footnote{
Poincar\'e-duality provides an isomorphism between the cohomologies 
for  pictures $(\pi^+, \pi^-)$ and $(-\pi^- -2, - \pi^+ -2)$ \cite{LZ2}.
Moreover, the cohomologies for pictures $(\pi^+ + \rho, \pi^- - \rho) $ 
with $\rho \in \frac{1}{2} \mathbb{Z}$ coincide due to spectral flow.
It is therefore sufficient to consider the case $\pi^{\pm} \geq -1$ only.}:
\begin{itemize}
\item 
The cohomology is empty for  pictures $(\pi^+, \pi^-) = 
\{ (-1,-1), (-1,0), (0,-1) \}$ and consists of the unit operator 
in the $(0,0)$ picture. 

\item 
In the  pictures $(-1,1)$ and $(1,-1)$ the cohomology consists 
of the spectral flow operators 
\begin{equation}\label{a} 
A(z) = (1 - c b') J^{--} e^{\varphi^+} e^{-\varphi^-} (z)
\end{equation}
and
$$
A^{-1} (z) =  (1 + c b') J^{++} e^{-\varphi^+} e^{\varphi^-} (z)
$$ 
with $J^{++} = \frac{1}{4} \epsilon_{ab} \psi^{+a} \psi^{+b}$ and 
$ J^{--} = - \frac{1}{4} \epsilon_{\bar{a}\bar{b}} {\psi}^{-\bar{a}}
{\psi}^{-\bar{b}}$ (see \cite{JLP} for conventions and a 
description of the $N{=}2$ string 
ghost system). One may check that $A$ and 
$A^{-1}$ are inverse to each other with respect to ring 
multiplication\footnote{Multiplication of two operators, denoted by a
dot in the following, means to take the regular term in  their 
operator product expansion \cite{W}.}.

\item 
In the  $(1,0)$ picture the cohomology consists of   the 
picture changing operator
$$ 
X^+ (z) = \{  Q , \xi^+ (z) \}
$$
{\it and} the operator 
$$
A \cdot X^- 
\hspace{1cm} {\rm with} \hspace{1cm}
X^- (z) = \{  Q , \xi^- (z) \}. 
$$ 
It should be emphasised that $A \cdot X^-$  is BRST inequivalent to $X^+$. 
Analogously, the $(0,1)$ cohomology 
consists of the operators $X^-$ and $A^{-1} \cdot X^+$.
\end{itemize}
We see that the size of the cohomology grows as the picture 
increases.
To obtain cohomology classes with higher integral
 picture numbers one may simply 
consider polynomials of  the operators $A$, $A^{-1}$ and $X^{\pm}$,
$$ 
\left ( X^+ \right )^k \cdot \left ( X^- \right )^\ell \cdot A^n, \hspace{1cm}
k, \ell \in \mathbb{N},\;\;\; n\in \mathbb{Z}.
$$ 
Note that $k$ and $\ell$ must not be negative since, 
contrary to $N{=}1$ strings, 
there do not exist local inverse picture changing 
operators for the $N{=}2$ string \cite{BKL} (the cohomology at 
vanishing momentum and ghost number is empty
for picture numbers $(-1,0)$ and $(0,-1)$).

It has been shown in \cite{JLP} that all these operators are 
{\it BRST inequivalent} ! For a given picture $(\pi^+,\pi^-)$
we thus have constructed $\pi^+ + \pi^- + 1$ operators,
\begin{equation}\label{coh} 
{\cal O}_{\pi^+,\pi^-, n}  =  (X^+)^{\pi^++n} \cdot (X^-)^{\pi^--n}\cdot A^n
, \hs n= - \pi^+ , ..., \pi^-.
\end{equation}
To obtain ghost number one cohomology classes of the closed string connected 
to the symmetries of the theory the operators in (\ref{coh}) 
have to be combined with right-moving cohomology classes of zero momentum 
and ghost number one. These operators can be found in a similar way: 
In \cite{Bi} it has been shown that the relevant cohomology in the $(0,0)$ 
picture is spanned by the four elements 
\begin{equation}\label{mom}
-i{\cal P}^a = c \partial Z^{a} - 2 \gamma^- \psi^{+a}, 
\hs -i \bar{\cal{P}}^{\bar a} = 
 c \partial \bar{Z}^{\bar a} - 2 \gamma^+ \psi^{-\bar a}.
\end{equation}
Here the target space Lorentz indices $a$ and $\bar a$ range from $0$ to $1$.
Multiplication with ${\cal O}_{\pi^+,\pi^-, n}$ gives similar operators in 
higher pictures: 
\begin{align}\label{coh1}
{\cal P}^a_{\pi^+,\pi^-, n} = {\cal O}_{\pi^+,\pi^-, n} 
\cdot {\cal P}^a, \nonumber \\
\bar{{\cal P}}^{\bar a}_{\pi^+,\pi^-, n} = {\cal O}_{\pi^+,\pi^-, n} \cdot 
\bar{{\cal P}}^{\bar a}.
\end{align}
We are now ready to write down the sought for closed string cohomology 
classes  of ghost number one: 
\begin{equation}\label{coh2}
{\Sigma}^{a}_{\pi^+,\pi^-,m,n} = {\cal O}_{\pi^+,\pi^-,m} (z)
\tilde{\cal P}^a_{\pi^+,\pi^-,n}(\bar z), \hs m,n = - \pi^+, ... ,\pi^-.
\end{equation} 
To save space the analogous operators $\Sigma^{\bar a}$ will not be
explicitly mentioned in the following.
Using the descent equations one may now construct an infinite set of 
symmetry charges and work out the transformation laws of the physical 
state. This has been done in \cite{JLP}. 

We conclude this section with one further remark. So far, we have only 
considered the relative cohomology of states that are  annihilated 
by the zero modes of all fermionic antighosts. It would, however, 
be more appropriate also to take into account states that are not 
annihilated by $b_0 + \tilde{b}_0$ which defines  the so-called semi-relative 
cohomology (one way to see that this is the right space to consider 
is to write down a kinetic term in a string field formalism). 
Allowing for more states generally changes the cohomology. But fortunately, 
one can show that the operators (\ref{coh2}) are still non-trivial 
in the semi-relative 
cohomology.   One may also wonder whether new  cohomology 
classes turn up, as happens  for  the bosonic string in two dimensions
\cite{WZ}. We do not know the general answer to this question, but explicit 
calculations for  low-lying pictures  indicate that this is not the case.    

\section{Ward identities}
We will now use the results from the previous section to derive 
Ward identities  for $N{=}2$ string amplitudes
 at arbitrary genus. Actually, an 
$N{=}2$ string scattering amplitude is further characterised by a Chern number
classifying  $U(1)$ bundles over the world-sheet  Riemann surface. It is, 
however, sufficient
to focus on vanishing  Chern number in the following. This will be justified 
in section four.
For reasons of space  the   
 general formalism will not be reviewed in detail here. Instead, we refer  
to  \cite{WZ, V,  KP}  for more extensive 
explanations. 

The basic object involved in the computation of scattering amplitudes  
is the vertex operator of the single degree of freedom
in the theory.  As usual, it splits into 
holomorphic and antiholomorphic parts: 
$$ 
V(z,\bar z, k) = V^{left} (z,k) \tilde{V}^{right} (\bar z , k)
$$ 
The left-moving operator is
$$ 
V^{left}_{(-1,-1)} (z,k) = cc' e^{-\varphi^+} e^{-\varphi^-} e^{ikZ^{left}}
$$ 
in the $(-1,-1)$ picture and 
$$ 
V_{(\pi^+, \pi^-)} (z,k)  = (X^+)^{\pi^++1} \cdot (X^-)^{\pi^-+1}
\cdot V_{(-1,-1)} (z,k) 
$$
in higher pictures (the right-moving piece $\tilde{V}^{right}$ looks 
similar)\footnote{Application of spectral flow only leads 
to vertex operators proportional to those above.}. 
Counting both metric and $U(1)$ but not supersymmetry ghost number 
vertex operators 
in closed $N{=}2$ string theory therefore have ghost 
number four (in our conventions picture changing operators have ghost number 
zero, see \cite{JLP}). Moreover, they are not annihilated by the zero 
modes $b'_0$ and $\tilde{b}'_0$ of the $U(1)$ antighosts. On the other hand
the ghost number one operators constructed in the previous section are 
all elements of the relative cohomology, i.e. they are all killed by the 
zero modes of all fermionic antighosts. It is, however, not too difficult 
to relate relative  cohomology classes to operators of higher ghost number, 
essentially by multiplying with the relevant ghosts. In this way we can 
construct from the ghost number one operators in equation (\ref{coh2}) 
new cohomology classes of ghost number three: 
$$ 
{\Sigma}^{a}_{\pi^+,\pi^-,m,n}  \to {\Omega}^{a}_{\pi^+,\pi^-,m,n}
\equiv c' {\tilde c}' \ {\Sigma}^{a}_{\pi^+,\pi^-,m,n} + \hdots 
$$ 
Here the  dots refer to further terms that might be necessary to achieve 
BRST invariance but  are unimportant otherwise. 

We are now ready to derive a Ward identity involving a genus $g$ scattering 
amplitude of $N$ external states with momenta 
$k_1, \hdots , k_N$ (denoted $A_N^g (k_1, \hdots , k_N)$ in the following). 
One starts with  the correlator\footnote{For simplicity we 
only consider closed string operators whose left- and right-moving picture 
numbers coincide.} 
\begin{equation}\label{corr} 
\left \langle 
\Omega^{a}_{\pi^+,\pi^-,m,n} (z, \bar{z} )  \,
\prod_{i=1}^N V_{\pi^+_i,\pi^-_i}^{cl}(z_i, \bar{z_i}, k_i)\, 
\prod_l (\mu_l , B ) (\tilde{\mu}_l , \tilde{B} ) 
\right \rangle_g
\end{equation}
where $(\mu_l , B )$ and $(\tilde{\mu}_l , \tilde{B} )$  are the  
the appropriate  Beltrami differentials integrated with 
the corresponding antighosts and the index $g$  
indicates that the correlator is meant to be evaluated 
with respect to the conformal field theory living on a Riemann surface of 
genus $g$. The antighosts can be applied to the vertex operators 
and the integrations can be pulled out of the brackets. Let us denote 
the remaining 
integrand by $\Theta$. If the operator $\Omega$ in (\ref{corr}) were replaced 
by an ordinary physical vertex operator $V$ one could integrate $\Theta$ 
over the moduli space of a genus $g$ surface with $N+1$ punctures. 
{}From  counting dimensions and ghost numbers it follows, however, that 
$\Theta$ as defined by 
(\ref{corr}) can be integrated only over the boundary of moduli space. In 
fact,  it can be considered as a differential form on moduli space 
of codimension one. Since $\Theta$   can also be shown to be a closed form 
\cite{WZ}  Stokes' theorem leads to the desired  Ward identity 
\begin{equation}\label{wi}
\int_{\partial {\cal M}^{g,N+1}} \Theta =  \int_{{\cal M}^{g,N+1}} 
d \Theta  = 0.
\end{equation}
The next step is to have a closer look at 
 the $N{=}2$ string moduli space ${\cal M}^{g, N+1}$, 
i.e. the moduli space  of  $N{=}2$ super Riemann surfaces 
with genus $g$ and  $N+1$ punctures (and vanishing Chern number in our case). 
In addition to    
the usual metric and super moduli, we also have to consider the so-called
$U(1)$ moduli \cite{BL2}  describing 
a continuum of possible monodromy phases for  the world-sheet fermions 
arising from their transport along
non-trivial homology cycles. However, the $U(1)$ moduli space 
is compact 
(it always has the topology of a torus) and therefore does not contribute 
to the boundary of moduli space. As a result, in our Ward identity (\ref{wi}) 
only the familiar boundary components of the metric moduli space appear.  

The metric moduli are of two different types. One  corresponds to  
the shape of the underlying Riemann surface whereas the other  
describes punctures, i.e. the locations of the vertex operators. 
If we move to the boundary of moduli space the Riemann surface degenerates in 
some way. In the following it is convenient to distinguish four
 different cases: 
First of all,  the underlying surface may pinch either along a trivial or 
a non-trivial homology cycle. If a genus $g$ surface pinches along a 
non-trivial cycle it becomes a surface of genus $g-1$ with two points 
coinciding. If it pinches 
along a trivial cycle the result is a connected pair of  Riemann surfaces 
with genera $g_1$ and $g_2$ such that $g_1 + g_2 = g$. 
For a $g{=}2$ surface with four punctures these two cases are 
illustrated in the top row of the figure below.
It may also happen that a number of  punctures approach each other. 
This is conformally equivalent to a situation where a sphere containing 
the relevant 
punctures  splits off of the rest of the surface. This is illustrated in
the bottom line of the figure, where we also distinguished whether 
two vertex operators $V$ approach each other or one $V$ approaches 
the ghost number three operator $\Omega$.

To see how a  pinch (denoted by $P$ in the figure) can properly be 
included 
in the computation let us recall that it can equivalently be 
described 
by an infinitely long cylinder. This cylinder can be taken into account 
by inserting a complete set of physical states.  In this formulation 
the twist angle of the cylinder is one of the moduli leading to 
an insertion of the metric  antighost combination $b(z)  - \tilde{b} 
(\bar z)$.
So the pinch can be represented by the sum 
\begin{equation}\label{csos}
\sum_i | \widehat{O}^i \rangle \langle O_i | 
\end{equation}
where $i$ labels a basis of the absolute BRST cohomology and 
\begin{equation}
\langle O_j |  O^i \rangle = \delta^i_j , \hs | \widehat{O}^i \rangle = 
(b_0 - \tilde{b}_0) |  O^i \rangle .
\end{equation}
What about  the fermionic and $U(1)$ moduli ? 
The former are correctly taken into account by obeying the right selection 
rules for picture numbers \cite{OL}.    
Moreover, a pinch contributes one complex $U(1)$ modulus. This corresponds 
to the fact that the complete set of states (\ref{csos}) carries two units of 
$U(1)$ ghost number -- just enough to compensate the antighost insertion 
due to the $U(1)$ modulus of the pinch.

Let us now  become more explicit: We assume $N\geq3$, i.e. the presence 
of at least three vertex operators,  and genus $g>0$
since tree-level 
amplitudes have been  discussed  in \cite{JLP}.
It will also  be sufficient and technically simpler 
to consider only  operators $\Sigma$ (and the corresponding $\Omega$) 
of the form 
$$ 
{\Sigma}^a_n (z, \bar z) := 
{\Sigma}^a_{-n,n,n,n} (z, \bar z) =
A^n (z)\,  
\tilde{A}^n \cdot {\tilde{\cal P}}^{a} (\bar z),
$$
which have picture numbers $(-n,n)$. The four 
cases mentioned above will now be discussed in turn.

\begin{figure}[ht]
\begin{center}
\epsfig{file=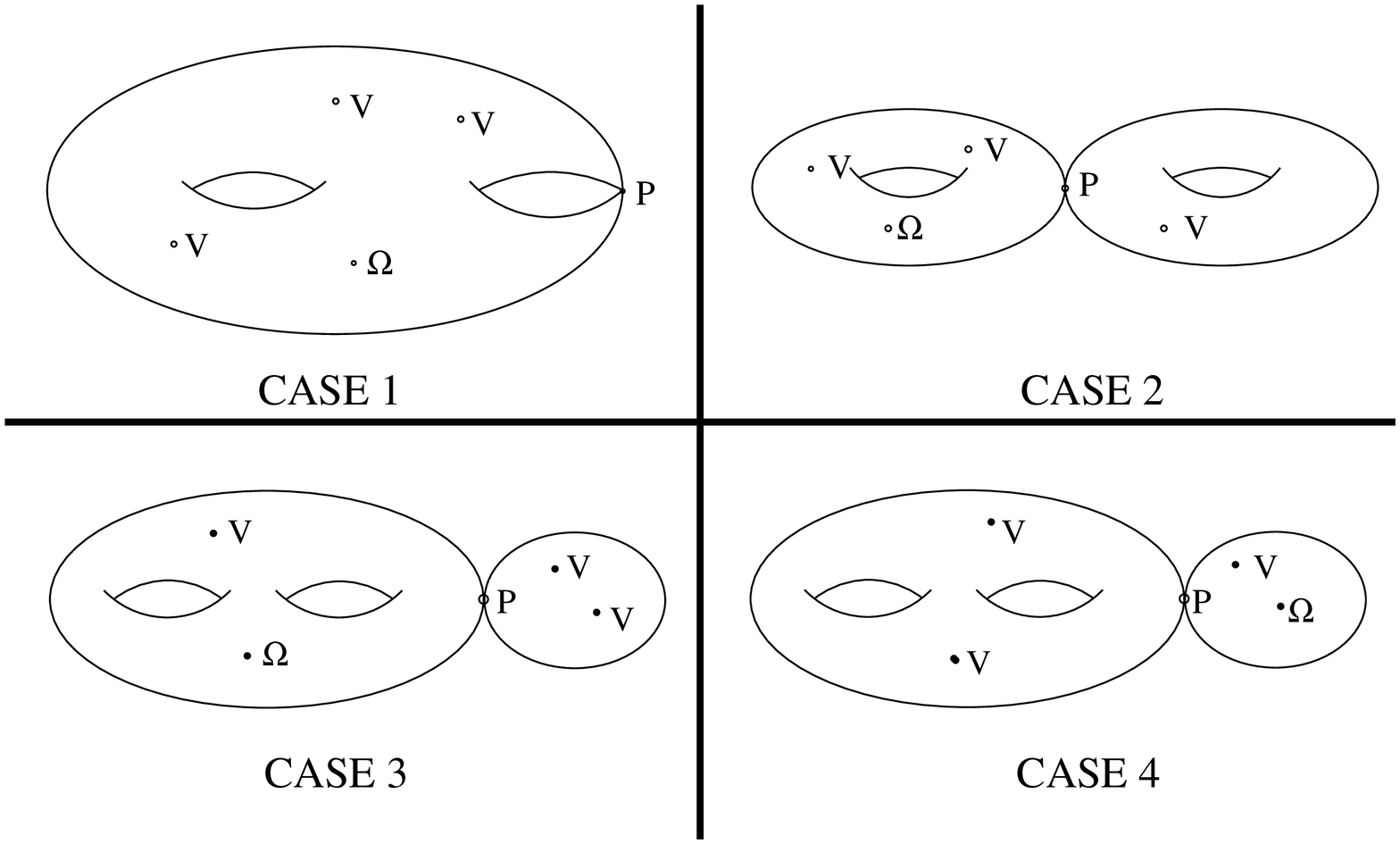, width=12cm}
\end{center}
\end{figure}

\subsection{Case 1: A non-trivial homology cycle pinches} 
Besides the $N$ physical vertex operators already present the pinching 
leads to an insertion of  two further  vertex operators $O_i$ and 
$\widehat{O}^i$, as explained above. So we have to evaluate 
the expression 
\begin{equation}\label{c1}
\sum_i \big\langle\!\big\langle \Omega_n^a\,  V_1 \, \hdots\,
 V_N \,   \widehat{O}^i\,  O_i  \big\rangle\!\big\rangle_{g-1}.
\end{equation}
Here, the notation for the vertex operators has been simplified in a
hopefully obvious way. The double bracket as usual denotes 
evaluation of the full amplitude including integration over moduli space. 

To further evaluate the expression (\ref{c1}) 
let us note that it contains at least 
six operators (since we assumed $N\geq 3$ in the beginning). 
Regardless of the value of $g$ integration 
over moduli space  leads for this number of operators  to
insertions of metric antighosts  that 
transform cohomology classes into integrated vertex operators.
Since this effect will be crucial in the following, 
we briefly review some details: \\
Assume the operator ${\cal B} (z) 
\tilde{\cal B} (\bar z)$ represents a closed string cohomology class. 
{}From the explicit form of the  BRST operator it follows that 
$$ 
{\cal B}^{(1)}  (z) = \oint_z \frac{dw}{2\pi i} b(w) {\cal B} (z) 
$$ 
satisfies the relation 
$$ 
\big [ Q , {\cal B}^{(1)} \big ] = \partial {\cal B},
$$ 
$Q$ being the left-moving part of the BRST operator. Since this argument 
goes through 
for the right-moving half, as well, a $b$-ghost insertion leads to the 
integrated operator
$$ 
\int d^2 z {\cal B}^{(1)}  (z) \tilde{\cal B}^{(1)} (\bar z)
$$ 
which is BRST invariant since the integrand transforms into a total 
derivative. In practice, going over from a cohomology class 
to an integrated vertex operator simply amounts to getting rid 
of the undifferentiated $c$- and $\tilde{c}$-ghosts. If some cohomology
class does not 
contain both these ghost fields (as for example the unit operator) its 
integrated form is zero.

We are always free to choose where to locate the  $b$-ghost 
insertions\footnote{Since we are dealing with vertex operators of 
non-standard ghost number, this is not completely obvious in  the 
path integral formulation.
In the operator formalism , however, one may explicitly check that the 
location of the $b$-ghost insertion is immaterial.}, 
i.e. which cohomology class to convert into an integrated operator. 
In the present case we can  pick $\Omega$. 
{}From the explicit form of $A$ in equation (\ref{a}) one sees that 
stripping off a $c$-ghost necessarily leads to the presence of a
$b'$-ghost, for example $A^{(1)} = - b' J^{--} e^{\varphi^+} e^{-\varphi^-}$. 
However,  there is no  corresponding $c'$-ghost in sight to compensate 
$b'$ in a correlation function. 
So we learn from simple
$U(1)$ ghost number counting that the amplitude (\ref{c1}) vanishes! 
In other words, the kind of degeneration considered in this subsection does
 not contribute to the Ward identity.

\subsection{Case 2: A trivial homology cycle pinches} 
The contribution to the Ward identity of  this component of the boundary 
is 
\begin{equation}\label{c2}
\sum_{i, \alpha}  \big\langle\! \big\langle V_{u_1} \hdots V_{u_p} 
\Omega_n^a\widehat{O}^i \big\rangle\! \big\rangle_{g_1}  
\big\langle\!\big\langle  O_i V_{u_{p+1}} 
\hdots V_{u_N} \big\rangle\!\big\rangle_{g_2}
\end{equation}
with $g_1 + g_2 = g$ and $g_1, g_2 > 0$. 
The sum over $\alpha$ runs over all possible ways to divide the set of 
$N$ physical 
vertex operators into a subset $\{ V_{u_1} \hdots V_{u_p} \}$ on the 
genus $g_1$
surface  and the remainder $\{ V_{u_{p+1}} \hdots V_{u_N}\} $ 
located  on  the other surface.

Since $g_1$ is strictly positive and the correlation function
 involving $\Omega$ 
contains at least one further operator the expression (\ref{c2}) can again be 
evaluated by transforming $\Omega$ to its integrated form. 
As in the previous subsection the vanishing of (\ref{c2}) then follows 
from $U(1)$  ghost number counting.
\subsection{Case 3: A sphere not including $\Omega$ splits off}
In this case we have to evaluate the expression 
\begin{equation}\label{c3}
\sum_{i, \alpha}  \big\langle\! \big\langle V_{u_1} \hdots V_{u_p} 
\Omega_n^a\widehat{O}^i \big\rangle\! \big\rangle_{g}  
\big\langle\!\big\langle  O_i V_{u_{p+1}} 
\hdots V_{u_N} \big\rangle\!\big\rangle_{g=0}.
\end{equation}
Since $g>0$ by assumption the correlator involving $\Omega$ vanishes by the 
same argument as above.
\subsection{Case 4: A sphere  including $\Omega$ splits off}
In this final case the contribution to the Ward identity reads
\begin{equation}\label{c4}
\sum_{i, \alpha}  \big\langle\! \big\langle V_{u_1} \hdots V_{u_p} 
\Omega_n^a\widehat{O}^i \big\rangle\! \big\rangle_{g=0}  
\big\langle\!\big\langle  O_i V_{u_{p+1}} 
\hdots V_{u_N} \big\rangle\!\big\rangle_{g}.
\end{equation}
The ghost number three operator $\Omega$  now appears in a tree-level 
amplitude whose evaluation involves metric antighost insertions 
as soon as more than three operators are present. 
Correspondingly, terms in the $\alpha$-sum vanish by the standard 
argument whenever the $g=0$ correlator involves 
 more than one operator $V$ besides $\Omega$ and $\widehat{O}^i$. 
What remains are those degenerations where $\Omega$ splits off with 
precisely one vertex operator $V$. These are the only contributions to the 
Ward identity:
\begin{equation}\label{wi2}
\sum_{u=1}^N  \sum_i \big\langle V_{u}  
\Omega_n^a \widehat{O}^i \big\rangle_{g=0}  
\big\langle\!\big\langle  O_i V_{1} \hdots V_{u-1}\, V_{u+1}
\hdots V_{N} \big\rangle\!\big\rangle_{g}= 0.
\end{equation}
Obviously,  the only non-vanishing term  in the above sum over $i$ occurs
 when $O_i$  coincides with the vertex operator $V_u$. In each term 
of the $u$-sum the second correlator 
therefore  is just the genus $g$ amplitude of $N$ physical states $A^g_N$. 
Reinserting the momenta $k_u$ allows us to rewrite the Ward identity as
\begin{equation}\label{wi3}
A_N^g (k_1 , \hdots , k_N ) \cdot \sum_{u=1}^N  \big\langle V (k_u)  
\Omega^a_n \widehat{V}(-k_u) \big\rangle_{g=0} = 0. 
\end{equation}
These  identities have already been  derived 
in \cite{JLP} for tree amplitudes. Equations (\ref{wi3}) tell us that 
they  do not get modified for  higher 
genera. The remaining  
correlator can be evaluated as 
\begin{equation}\label{hk} 
\big\langle V (k)  
\Omega_n^a \widehat{V}(-k) \big\rangle_{g=0} = \left ( \frac{\bar{k}^0}{k^1} 
\right )^n k^a \equiv h(k)^n k^a.
\end{equation}
The final identities  for the genus $g$ amplitude thus read 
\begin{equation}\label{fwi}
A_N^g (k_1 , \hdots , k_N ) \cdot \sum_{i=1}^N h(k_i)^n k_i^a = 0
\hspace{1cm}
\mbox{for any}
\hspace{.2cm}
 n \in  \mathbb{Z}
\end{equation}
and imply the vanishing of all amplitudes with $N\geq 4$ \cite{JLP}.
The three point function, however, is generally non-zero. 
One may for example check that the tree-level amplitude 
$$
A^{g=0}_{N=3} (k_1, k_2, k_3) = \big ( {\bar k}_1 \cdot k_2 - 
{\bar k}_2 \cdot k_1 \big )^2 
$$ 
satisfies all identities without being zero. 
On dimensional grounds  it seems very plausible that for higher genus 
the three point function is just a power of the tree-level result: 
$$
A^{g}_{N=3} (k_1, k_2, k_3) = \alpha_g \big ( {\bar k}_1 \cdot k_2 - 
{\bar k}_2 \cdot k_1 \big )^{4g+2} 
$$ 
Here the pre-factor $\alpha_g$ depends on the genus but not on the momenta. 
Explicit computations at one loop show that $\alpha_{g=1}$ is divergent 
\cite{BGI, N}.
This concludes our discussion of the scattering amplitudes of the $N{=}2$ 
string.

\section{Some remarks}
So far we have ignored the possibility of non-vanishing Chern number $c$,  
 corresponding to topologically non-trivial configurations of the 
$U(1)$ gauge field on the world-sheet. A careful evaluation of the 
path integral  shows that a non-zero Chern number can be simulated  by 
inserting (a power of) the spectral flow operator $A$ into the 
$c=0$  correlation function and simultaneously adjusting the picture numbers 
of the vertex operators \cite{BL2}. Since 
the derivative of the spectral flow  operator  is BRST trivial each 
$A$ (or $A^{-1}$)  can be moved towards  
one of the vertex operators and simply pulls out a momentum factor $h(k)$ 
(or its inverse, see  eq. (\ref{hk}) for a definition of $h$). 
Therefore, amplitudes with different Chern number 
are proportional to one another. 
Hence,  it is sufficient to prove the vanishing of 
a scattering amplitude for one fixed value of $c$. 
Secondly we have ignored that, as a Riemann surface with $c=0$ 
 degenerates and splits into two, the resulting surfaces may have 
non-vanishing Chern numbers $c$ and $-c$. So we actually should include in our 
Ward identity a summation over all such splittings\footnote{This 
sum is finite since supersymmetry ghost zero modes kill correlators 
when $|c|$ exceeds a certain value.}. However, we have just    
explained that this only leads to additional factors $h(k)^c$ and $h(k)^{-c}$
which cancel each other ($k$ is the momentum flowing through the pinch). 
This justifies our treatment where we completely 
neglected sectors with non-zero Chern number.

A further point that deserves to be mentioned is the question of non-linear 
contributions to the symmetries. One of the remarkable features of the 
$N{=}2$ string that make it such an interesting toy model
is the fact that we know a simple  field theory that reproduces 
the  tree-level amplitudes to all orders in $\alpha'$. 
This field theory is well known to possess a highly non-linear 
symmetry structure. In \cite{JLP} the linearised version of the unbroken 
symmetries on the field theory side was compared to the 
transformation rules of the $N{=}2$ string vertex operators under the 
symmetries that lead to the above Ward identities. They were found to 
coincide. In fact, the Hilbert space in our formulation of the theory 
consists only of  single string states.  So it seems at first sight correct 
to restrict a comparison between symmetries in field theory and string theory 
to the linear level. However, it has been explained in \cite{WZ} (section 6) 
that non-linear symmetry structures can make their appearance in a first 
quantised string theory at the level of Ward identities. More precisely, 
a non-linear contribution to a Ward identity corresponds to a situation 
where the  ghost number one (three for $N{=}2$ strings) 
operator $\Omega$ splits off with more than one further vertex operator. 

\begin{figure}[ht]
\begin{center}
\epsfig{file=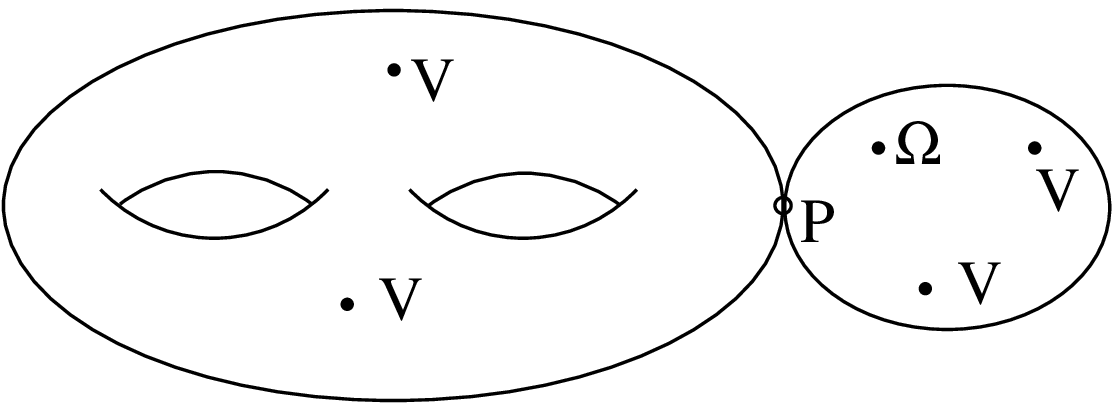, width=8cm}
\center{Non-linear contribution to the Ward identity}
\end{center}
\end{figure}
\noindent
In this case only the overlap between the charge acting on a single  vertex 
operator with a multi-string state  is sent through the pinch. 
In other words, a symmetry is realized non-linearly precisely when 
the tree-level amplitude 
$$
\big\langle\! \big\langle \Omega\, V (k_1) \hdots V (k_{n-1}) 
\widehat{V} (k_n) \big\rangle\! \big\rangle_{g=0} 
$$ 
is non-vanishing for $n\geq 3$. A model where this indeed happens is the 
bosonic string in two dimensions. Yet it has been argued in section three 
that in our case of the $N{=}2$  string the relevant correlation functions 
vanish. As a consequence the Ward identity (\ref{fwi}) is linear.
This indicates a clear discrepancy 
to the field theory  and suggests that the behaviour of the $N{=}2$ string
is not fully captured by its tree-level effective field theory. 

Last but not least we should give our opinion on  the reliability 
of our arguments. In fact, we must admit that the analysis 
of  the boundary of the $N{=}2$ string moduli space has been somewhat 
heuristic. 
It is mainly based on counting of dimensions and ghost numbers. 
Hidden  subtleties might be detected by a more careful 
investigation. 
For example, it is conceivable that the $U(1)$ moduli space behaves in 
some discontinuous way as the Riemann surface degenerates. 
Whether or not this is the case can only be answered by studying the 
relevant index theorem. Other potential difficulties are related to the 
fermionic moduli that we have treated 
in a rather straightforward way, ignoring possible  ambiguities  due to 
the location of picture changing operators. In any case it would be  
helpful to have an explicit computation of the one-loop four point function. 
If that turns out to be non-vanishing it will be extremely interesting 
to see by which mechanism the derivation of the Ward identities must be 
modified.

\end{document}